\documentstyle[12pt]{article}

 \def\void{}
\def\labelmark{}

\newenvironment{formula}[1]{\def\labelname{#1}
\ifx\void\labelname\def\junk{\begin{displaymath}}
\else\def\junk{\begin{equation}\label{\labelname}}\fi\junk}%
{\ifx\void\labelname\def\junk{\end{displaymath}}
\else\def\junk{\end{equation}}\fi\junk\labelmark\def\labelname{}}

{\ifx\void\labelname\def\junk{\end{array}\end{displaymath}}
\else\def\junk{\end{array}\right.\end{equation}}
\fi\junk\labelmark\def\labelname{}\def\junk{}
}

\newcommand{\beq}{\begin{formula}}
\newcommand{\eeq}{\end{formula}}
\newcommand{\beqv}{\begin{formula}{}}

\newcommand{\rf}[1]{(\ref{#1})}
\newcommand{\oh}{\frac{1}{2}}

\newcommand{\bea}{\begin{eqnarray}}
\newcommand{\eea}{\end{eqnarray}}
\newcommand{\beas}{\begin{eqnarray*}}
\newcommand{\eeas}{\end{eqnarray*}}
\newcommand{\beqs}{\begin{displaymath}}
\newcommand{\eeqs}{\end{displaymath}}

\newcommand{\br}{\langle}
\newcommand{\kt}{\rangle}

\renewcommand{\a}{\alpha}

\newcommand{\vp}{\varphi}

\newcommand{\cD}{{\cal D}}

\newcommand{\cH}{{\cal H}}

\newcommand{\Tr}{{\rm Tr}\;}
\newcommand{\ben}{\begin{equation}}
\newcommand{\een}{\end{equation}}

\newcommand{\bdm}{\begin{displaymath}}
\newcommand{\edm}{\end{displaymath}}

\newcommand{\pa}{\partial}

\newcommand{\bbR}{{\bf R}}

\begin{document}
 \topmargin 0pt
 \oddsidemargin 5mm
 \headheight 0pt
 \topskip 0mm

 \addtolength{\baselineskip}{0.4\baselineskip}

 \pagestyle{empty}

 \vspace{0.1cm}

\hfill  

\hfill

\vspace{1cm}

\begin{center}

{\Large \bf Noncommutative Waves have}

\medskip

{\Large\bf Infinite Propagation Speed}



\vspace{1.2 truecm}



 \vspace{0.7 truecm}
{\bf Bergfinnur Durhuus}\footnote{e-mail: durhuus@math.ku.dk}

\vspace{0.5 truecm}

Matematisk Institut and MaPhySto\footnote{MaPhySto - A network in Mathematical Physics and Stochastics, funded by the Danish National Research Foundation.}, Universitetsparken 5

2100 Copenhagen \O, Denmark

 \vspace{.8 truecm}

{\bf Thordur Jonsson}\footnote{e-mail:
thjons@raunvis.hi.is}\footnote{Permanent address: University of
Iceland, Dunhaga 3, 107 Reykjavik, Iceland}

\vspace{0.5 truecm}

Matematisk Institut, Universitetsparken 5

and

The Niels Bohr Institute, Blegdamsvej 17

2100 Copenhagen \O, Denmark


 \vspace{1.5 truecm}

 \end{center}

 \noindent
 {\bf Abstract.} We prove the existence of global solutions to the
Cauchy problem for noncommutative 
nonlinear wave equations in arbitrary even spatial dimensions where the 
noncommutativity is only in the spatial directions.  We find that 
for existence there are
no conditions on the degree of the nonlinearity 
provided the potential is positive.
We furthermore prove that nonlinear noncommutative waves 
have infinite propagation speed, i.e.,
if the initial conditions at time $0$ have a compact support then for any
positive time the support of the solution can be arbitrarily large.

 \vfill

 \newpage
 \pagestyle{plain}

\section{Introduction}  
Classical solutions to the static equations of field theories
in noncommutative spaces have been studied in some detail in recent years.
Most of these theories are defined in the Moyal plane, its higher 
dimensional generalizations or on the fuzzy sphere.  
Explicit solitonic solutions 
have been found in various gauge theories, see, e.g.,
\cite{g1,g2,g3,g4,g5} 
as well as in 
scalar field theories \cite{h1,h2,h3} at infinite noncommutativity  
where the existence theory for finite noncommutativity 
is now rather complete \cite{p1,p2,p3,p4} especially for the rotationally
invariant case.  For general
background and reviews of
noncommutative field theory we refer to \cite{r1,r2,r3}. 

Some time dependent solutions to noncommutative field equations 
have been found in gauge theories and sigma models, see, e.g., 
\cite{bak,td1,td2,td3, popov1,popov2,bieling}, 
and also by boosting static solutions in scalar 
field theories \cite{boost}.
The main purpose of this paper is to study the initial value problem for 
nonlinear wave equations in odd dimensional noncommutative Lorentzian 
spaces with 
the noncommutativity in the spatial directions only.
We prove the existence  
of global solutions for a large class of initial conditions provided 
the nonlinear term in the wave equation is the derivative of a positive
polynomial which vanishes to second order at zero.
We do not need to impose any conditions on the degree of the nonlinearity. 
This is in stark contrast with the classical case 
where the existence theory depends strongly on the dimension of space 
as well as the nonlinearity \cite{strauss}.

Our existence results are obtained by adapting the standard theory of
evolution
equations in Banach spaces, see, e.g., \cite{reed,reedsimon} and
references therein, to
the setting of  noncommutative wave equations.
The arguments presented in this paper can be generalized to 
nonpolynomial nonlinearities, 
e.g., entire functions.

We analyse the support properties of the solutions and show 
quite generally that the support is arbitrarily large 
in all spatial slices at 
positive times even though the initial data at time $0$ have compact 
supports.  Again
this is qualitatively different from classical nonlinear waves which
always have a propagation speed equal to that of the corresponding
linear equation \cite{strauss}. On the other hand, it should be noted that the noncommutative wave equation under consideration is not Lorentz invariant, since a specific choice of spatial slicing of Minkowski space is assumed. 

\section{Existence of Solutions}
Let $\nabla^2$ denote the Laplacian in $\bbR^{2d}$.  For simplicity we
choose the nonlinear term in the wave equation to be a polynomial $F$.
We assume $F(s)=V'(s)$ where $V(s) > 0$ for all
$s\neq 0$,
$V(0)=V'(0)=0$ but
$V''(0) >0$. 
We study wave equations of the form
\beq{1}
(\pa_t^2-\nabla^2)\vp (t,x)+F_*(\vp )(t,x)=0,
\eeq
where $F_*(\vp )$ is defined by replacing the ordinary pointwise product
of functions by the standard star product for functions on $\bbR^{2d}$, i.e.,
\beq{2}
F_*(\vp )=\sum_{n=1}^{N}c_n\, \vp^{*n}\,,
\eeq
if $F(s)=\sum_{n=1}^{N}c_n\, s^n$ and $\vp^{*n}$ denotes the $n$'th $*$-power of $\vp$. 
The star product of two functions $f$ and $g$ is defined by
\beq{2x}\left.
(f*g)(t,x)=\exp\left(-\frac{i}{2}\Theta_{ij}{\pa\over\pa y_i}
{\pa \over\pa z_j}\right)f(t,y)g(t,z)\right|_{y=z=x},
\eeq
where $\Theta$ is a nondegenerate antisymmetric $2d\times 2d$ matrix.
A more enlightening formula for the star product, which clearly
exhibits its nonlocal character, is given by 
\beq{2xx}
(f*g)(t,x)={1\over (2\pi)^{2d}}\int f(t, x+\oh \Theta p)g(t, x+y)
e^{ip\cdot y}\,dpdy
\eeq
which can be obtained from \rf{2x} by Fourier transforming.

We are interested in real valued solutions $\vp$ to Eq.\ \rf{1}.
We will assume for simplicity that the
noncommutativity matrix $\Theta$ is a direct 
sum of $d$ $2\times 2$ antisymmetric matrices 
$$
\left(\begin{array}{rr}0&1\\-1&0\end{array}\right),
$$
times a positive noncommutativity
parameter, denoted $\theta$. Our results are, however, valid for arbitrary non-degenerate antisymmetric $\Theta$.

It is convenient to apply the Weyl 
quantization map to the wave equation and then, with our conventions, 
it takes the form
\beq{3}
\pa_t^2\phi (t)+2\theta^{-1}\sum_{k=1}^d[a^*_k,[a_k,\phi (t)]]+F(\phi(t) )=0,
\eeq
where $\phi (t)$ is a self-adjoint operator on $L^2(\bbR^d)$ and the operators 
$a_k^*$ and $a_k$ are $d$ independent pairs of raising and lowering operators 
whose explicit form can be taken to be  
\beq{4}
a_k={1\over\sqrt{2}}\left(\xi_k+{\pa\over\pa\xi_k}\right),
\eeq
where $\xi_k$, $k=1,\ldots ,d$, are the coordinates in $\bbR^d$.
The relation between a function $\vp$ on $\bbR ^{2d}$ 
and the corresponding Weyl operator
$\pi (\vp )= \phi$ on $L^2(\bbR ^d)$ is given by
\beq{5}
\phi ={1\over (2\pi)^d}\int \tilde{\vp}(w_1,\ldots ,w_{2d})
\exp\left(i\sum_{j=1}^d (w_{2j-1}\hat{\xi}_j + w_{2j}\hat{p}_j)\right)
\, dw,
\eeq
where $\hat{\xi_k}$ and $\hat{p}_k$ are the position and momentum
operators corresponding to the raising and lowering operators $a^*_k$
and $a_k$ and $\tilde{\vp }$ denotes the Fourier transform of $\vp$.  
The Weyl map is in fact an algebra isomorphism from $L^2(\bbR
^{2d})$ (with the star product) 
to the space of Hilbert-Schmidt operators on $L^2(\bbR ^d)$ and takes
real valued functions to self-adjoint operators.

The collection of all Hilbert-Schmidt operators on $L^2(\bbR ^d)$ 
forms a Hilbert space, denoted 
$\cH ^2$, with inner product
\beq{6}
\br A,B\kt _2=\Tr (A^*B).
\eeq
We define a self-adjoint operator $\Delta$ on $\cH ^2$ by
\beq{6y}
\Delta\pi (\vp )=\pi (\nabla^2\vp )
\eeq
and it is easily checked that
\beq{6yy}
\Delta\phi =-2\sum_{k=1}^d[a^*_k,[a_k,\phi ]].
\eeq
Its domain is ${\cal D}(\Delta)=\{\pi(\vp):\;\vp,\nabla^2\vp\in L^2(\bbR^d)\} $.

It is convenient to write the wave equation as two first order
equations in $t$:
\bea
\pa_t\phi_1 & = & \phi_2\\
\pa_t\phi_2 & = & \Delta\phi_1-\theta F(\phi_1).
\eea
Here we have also rescaled the time variable by $\theta^{-1/2}$.  Writing
\beq{6xxzz}
\Phi =\left(\begin{array}{cc}\phi_1\\ \phi_2\end{array}\right),
\eeq
for ordered pairs of operators, and defining
\beq{7}
A=i\left(\begin{array}{cc}0 & I\\ \Delta & 0\end{array}\right),~~~~
J\left(\begin{array}{cc}\phi_1\\  \phi_2\end{array}\right)=
\left(\begin{array}{cc}0\\- F(\phi_1)\end{array}\right),
\eeq
we can write the wave equation on the form
\beq{8}
\pa_t \Phi =-iA\Phi +\theta J(\Phi),
\eeq
where $A$ is a linear operator acting on pairs of operators.

We now prove existence of solutions to \rf{8} using standard mehods
from the theory of nonlinear evolution equations which can also be
used to deal with ordinary nonlinear wave equations, see
\cite{reed,reedsimon}.  
First let us note
that we can replace the operator $\Delta$ in \rf{8} by $\Delta -m^2$,
$m>0$, by adding a compensating linear term to $F$ and the modified $F$ 
is still the derivative of a positive polynomial, provided $m$ is small enough.   Define a self-adjoint 
operator $B$ on $\cH ^2$ by
\beq{9}
B=(-\Delta +m^2)^{1/2}.
\eeq
Then $B\geq m$ and we define the space $\cH ^{1,2}$ as the
set of operators $\phi\in\cH ^2$ for which $B\phi\in\cH ^2$.
Let $\cH=\cH ^{1,2}\oplus\cH ^2$  and define an inner product 
$\langle\cdot\, , \cdot\rangle$ on $\cH$ by
\beq{10}
\langle\left(\begin{array}{cc}\phi_1\\ \phi_2\end{array}\right) ,
\left(\begin{array}{cc}\psi_1\\ \psi_2\end{array}\right)\rangle =
\br B\phi_1, B\psi_1\kt _2 +\br \phi_2 ,\psi _2\kt _2.
\eeq
With this inner product $\cH$ is a Hilbert space and the operator
\beq{33}
D=i\left(\begin{array}{cc}0 & I\\ -B^2 & 0\end{array}\right)
\eeq
is a self-adjoint operator on $\cH$ with domain ${\cal D}(D)= {\cal D}(\Delta)\oplus \cH^2$. We can now apply the following theorem. 

\medskip   
\noindent
{\bf Theorem.}  {\it Let $T$ be a self-adjoint operator with domain
$\cD (T)$ on a Hilbert space $H$ with norm $\|\cdot \|$ and $N:\cD
(T)\mapsto \cD (T)$ a mapping such that the following conditions hold
for any $\phi, \psi\in\cD (T)$:
\begin{enumerate}

\item[(i)] $\| N(\phi )\|\leq C_1(\|\phi\| )\|\phi \|$

\item[(ii)] $\| TN(\phi )\|\leq C_2(\|\phi\| )\|T\phi \|$

\item[(iii)] $\|N(\phi )-N(\psi )\|\leq C_3(\|\phi \|, \|\psi\|) \|\phi
-\psi\|$

\item[(iv)] $\|T(N(\phi )-N(\psi ))\|\leq C_4(\|\phi \|, \|\psi\|,
\|T\phi\| , \| T\psi\|) \|T\phi
-T\psi\|$

\end{enumerate}
where the constants $C_j$, $j=1,2,3,4$, are increasing functions of
the norms.  Then, for any $\phi_0\in\cD (T)$ there exists $t_0>0$ and
a unique
continuously differentiable family of vectors $\{\phi (t)\}_{0\leq t<
t_0}\subseteq\cD (T)$ such that
\beq{13}
\pa_t\phi (t)=-iT\phi (t) + N(\phi (t))
\eeq
and
\beq{14}
\phi (0)=\phi_0.
\eeq
If $\|\phi (t)\|$ is bounded by a constant independent of $t$ then one can
take $t_0=\infty$. 

}
\medskip

A proof of this theorem can be found in \cite{reedsimon}.  The basic
step is to convert the differential equation (together with the
initial conditions) into the integral equation
\beq{15}
\phi(t)=e^{-itT}\phi_0+\int_0^te^{-i(t-s)T}N(\phi (t))\,ds.
\eeq
It is quite straightforward to verify conditions $(i)$-$(iv)$
for $N=\theta J$, $T=D$ and the Hilbert space $\cH$ that we defined before.
The uniform bound on the norm of the solution is obtained from the
conservation of energy.

For illustration we verify $(i)$ and 
$(iv)$ leaving the others, which
are similar,  to the reader.
Let us denote the norm on the space of Hilbert-Schmidt operators by
$\|\cdot \|_2$.  Then we have for the norm $\|\cdot \|$ on $\cH$
\beq{15x}
\|\left(\begin{array}{cc}\phi_1\\ \phi_2\end{array}\right)\|^2=
\|B\phi_1\|_2^2+\|\phi_2\|_2^2
\eeq
and clearly
\beq{16}
\|J\left(\left(\begin{array}{cc}\phi_1\\
\phi_2\end{array}\right)\right)\|^2 =\|F(\phi_1)\|_2^2.
\eeq
For $(i)$ we therefore need to show that
\beq{17}
\|F(\phi_1)\|_2\leq C_1 (\|B\phi_1\|_2)\|B\phi_1\|_2.
\eeq
Since $F$ is a polynomial it suffices to verify the inequality \rf{17}
for the case $F(s)=s^n$ and this follows from 
\bea
\|\phi_1^n\|_2 & \leq & \|\phi_1\|^n_2\nonumber\\
               & \leq & \left(m^{-n}\| B\phi_1\|^{n-1}_2\right)
               \| B\phi_1\|_2.
\eea

For condition $(iv)$ it is again enough to consider the case when $F$ is
a monomial.   With $F$ as before we have
\beq{18}
\| DJ\left(\left(\begin{array}{cc}\phi_1\\
\phi_2\end{array}\right)\right)-DJ\left(\left(\begin{array}{cc}\psi_1\\
\psi_2\end{array}\right)\right)\|=\|B\phi_1^n-B\psi_1^n\|_2
\eeq
and
\beq{19}
\| D\left(\begin{array}{cc}\phi_1\\
\phi_2\end{array}\right)-D\left(\begin{array}{cc}\psi_1\\
\psi_2\end{array}\right)\|^2 = \|B(\phi_2-\psi_2)\|_2^2 +
\|B^2(\phi_1-\psi_1)\|_2^2
\eeq
so we need to show that
\beq{20}
\|B(\phi_1^n-\psi_1^n)\|_2\leq C \; \|B^2(\phi_1-\psi_1)\|_2,
\eeq
where the constant $C$ is an increasing function of the norms as
stated in the Theorem.
Let us define operators $A_k$ on $\cH ^{1,2}$ by
\beq{21}
A_k\phi=[a_k,\phi],
\eeq
for $k=1,\ldots , d$.  The operators $A_k$ are derivations and 
\beq{22}
-B^2+m^2=\Delta =-2\sum_{k=1}^d A_k^*A_k.
\eeq
We can write
\beq{23}
\phi_1^n-\psi_1^n=\sum_{i=0}^{n-1}\phi_1^{n-1-i}(\phi_1-\psi_1)\psi_1^i
\eeq
so in order to establish \rf{20} it suffices to estimate
\beq{24}
\|B(\phi_1^i(\phi_1-\psi_1)\psi_1^j)\|_2^2  = 
m^2\|\phi_1^i(\phi_1-\psi_1)\psi_1^j)\|_2^2 +2\sum_{k=1}^d
\|A_k(\phi_1^i(\phi_1-\psi_1)\psi_1^j)\|_2^2
\eeq
with $i+j=n-1$.
The first term on the right hand side above is bounded by
$$
m^{-2} \|\phi_1\|_2^{2i}\|\psi_1\|_2^{2j}\|B^2(\phi_1-\psi_1)\|_2^2
$$
so it suffices to consider the second term.  We have
\bea
\lefteqn{\|A_k(\phi_1^i(\phi_1-\psi_1)\psi_1^j)\|_2  \leq 
i\,\|A_k\phi_1\|\,\|\phi_1\|^{i-1}\|\phi_1-\psi_1\|_2\|\psi_1\|^j }
\nonumber\\ &&
 +\|\phi_1\|^{i}\|A_k(\phi_1-\psi_1)\|_2\|\psi_1\|^j
+j\,\|\phi_1\|^{i}\|\phi_1-\psi_1\|_2\|\,\|A_k\psi_1\|\,\|\psi_1\|^{j-1}. 
\eea
Since 
\beq{26}
\|A_k\chi\|_2\leq \|B\chi\|_2
\eeq
for $k=1,\ldots ,d$ and any $\chi$ in the 
domain of $B$ the desired result follows easily by
summing over $k$.

It follows from the differentiability of the solution 
\beq{e0}
\Phi (t)=\left(\begin{array}{cc}\phi_1 (t)\\
\pa_t\phi_1(t) \end{array}\right)
\eeq
and the equation of
motion that the energy
\beq{e1}
E(t)={\rm Tr}\left( \sum_{k=1}^d [a_k,\phi_1 (t)][\phi_1 (t),a^*_k]+\oh 
(\pa_t\phi_1 (t))^2+\theta V(\phi_1 (t))\right)
\eeq
is independent of $t$. Since $V(x)\geq\oh m^2x^2$ for real $x$ it 
follows that, for $\varepsilon >0$ sufficiently small, we have
\beq{e3}
E(t)\geq \varepsilon \|\Phi (t) \|^2 
\eeq
provided $\Phi(t)$ is self-adjoint, which is the case if the initial conditions are self-adjoint.
Thus, we have proven that {\it all solutions to Eq.(\ref{1}) with real initial 
data $\vp (0,\cdot)\in {\cal D}(\nabla^2)$ and $\partial_t\vp(0,\cdot)\in L^2(\bbR^{2d})$ are global.}

We remark that the estimates above for the noncommutative case are in general
not valid for the corresponding classical (commutative) nonlinear equations.  
For example, the Sobolev inequality
needed to establish the conditions {\it (i)-(iv)} in the classical case is 
valid for arbitrary polynomials $F$ if $d=1$, whereas for $d\geq 2$ it only holds for linear $F$. 
Star powers of functions are
more regular than ordinary powers of functions.  Roughly speaking, the local 
singularities of the star product of two functions are no worse than the
singularities of the individual functions.   We will discuss this in
more detail in the following section.  It follows from this
enhanced regularity that the existence
problem for nonommutative nonlinear waves is mathematically similar to the
existence problem for nonlinear waves in the case where space has been
replaced by a discrete lattice.  

\section{Infinite Propagation Speed}
In this section we show that in general 
the diameter of the support of a solution 
to the
noncommutative wave equation increases in time with
infinite speed.   By support we here mean the support of the function
which corresponds to the operator under the Weyl map.  

We first note that the solutions whose existence was established in
the preceding section can be written
\beq{28}
\phi (t)= (\cos tB)\phi (0) +(B^{-1}\sin tB )\pa_t\phi (0)-
\theta\int_0^t B^{-1}\sin((t-s)B)F(\phi (s))\,ds
\eeq
by using \rf{15} and the expression
\beq{29}
e^{-itD}=\left(\begin{array}{cc}\cos tB & B^{-1}\sin tB\\
-B\sin tB & \cos tB \end{array}\right) 
\eeq
for the linear time development operator acting on $\cH$.
Let $\vp(t,\cdot )$ be the function corresponding to $\phi (t)$ under the Weyl
map, let $g(t,\cdot )$ correspond to
\beq{30}
\chi (t) = (\cos tB)\phi (0) +(B^{-1}\sin tB )\pa_t \phi (0)
\eeq
and let $h(t,\cdot )$ correspond to
\beq{31}
\psi (t)= \theta\int_0^t B^{-1}\sin((t-s)B)F(\phi (s))\,ds .
\eeq
Then $g$ is a solution to the linear wave equation and 
the support of $g$ is contained in the ordinary causal future of
the union of the supports of $\vp(0, \cdot )$ and $\pa_t\vp(0, \cdot )$ by
the Huygen's principle for solutions to linear wave equations.    We are
therefore interested in studying the support properties of $h(t, \cdot )$.

We note that $\phi (t)\to \phi_0$ and hence also $F(\phi (t))\to
F(\phi_0)$ as $t\to 0$ where the convergence is in Hilbert-Schmidt norm.  
Since the operator
$ (tB)^{-1}\sin(t B)$ converges strongly to the identity operator $I$ 
as $t\to 0$ and is uniformly bounded in norm it follows
from Eq.\ \rf{31} that
\beq{32}
\| \psi (t)-\oh t^2 \theta F(\phi_0)\|_2 = o(t^2)
\eeq
as $t\to 0$.   It follows that the relevant support properties of $h(t,\cdot
)$ for small $t$ are determined by those of $F_*(\vp_0)$ where $\phi_0=\pi (\vp_0)$.
Suppose that $\vp_0$ is smooth with a compact support $S_0$ and assume that we
can show that the support of $F_*(\vp_0)$ is strictly larger
than $S_0$ in the sense that there is a closed ball $U$ disjoint from $S_0$ with
\beq{32xx}
\int_U |F_*(\vp_0)|^2\,dx\equiv A>0.
\eeq
Then, by Eq.\ \rf{32}, we have
\beq{32xy}
\int_U |h(t,x)|^2\, dx\geq \oh t^2\theta A+ o(t^2)
\eeq
and $h(t,\cdot )$ is nonzero on a subset of $U$ of positive measure for all 
$t$ small enough.  Evidently
this proves that the propagation speed is infinite for the initial values 
$\vp(0,\cdot )=\vp_0$ and $\pa_t\vp(0,\cdot )$ smooth with support in $S_0$.

It remains to demonstrate that the support of $F_*(\vp_0)$ is in general strictly larger than that of $\vp_0$. 
The support properties of star products of functions are 
simply reflected in the
algebra of star products of $\delta$-functions and their Fourier transforms.
We define the distributions $D(\alpha )$, $E(\beta )$ for $\alpha , \beta\in
\bbR^{2d}$ by
\beq{32xyz}
D(\alpha)=\pi^d \delta_\a~~{\rm and}~~E(\beta )=e^{-2i\Theta (\beta ,\cdot )}
\eeq
where $\delta_\alpha $ is delta function supported at $\a$ and
$\Theta$ is the skew symmetric quadratic form on $\bbR ^{2d}$ entering 
Eq.\ (\ref{2x})
\beq{uuu}
\Theta (\alpha,\beta)=\Theta_{ij}\alpha_i\beta_j\;.
\eeq
  By a straightforward computation, using \rf{2xx}, one
finds
\bea
D(\alpha )*D(\beta ) & = & e^{2i\Theta (\alpha ,\beta)} E(\alpha -\beta )\label{a1} \\
D(\alpha )*E(\beta ) & = &  e^{2i\Theta (\alpha ,\beta)} D(\alpha -\beta )  \label{a2} \\
E(\alpha )*D(\beta ) & = & e^{-2i\Theta (\alpha ,\beta)} D(\alpha +\beta )  \label{a3}  \\
E(\alpha )*E(\beta ) & = &  e^{-2i\Theta (\alpha ,\beta)} E(\alpha +\beta ).\label{a4}
\eea
The Weyl map can in fact be extended to distributions using \rf{2xx}
and one can show that the star product of two tempered distributions
is again a tempered distribution provided they are sufficiently regular or 
one of the two distributions has a compact support.

We can write
\beq{40}
f=\int f(\alpha )\delta_{\alpha }\,d\alpha 
\eeq
for a function $f$ on $\bbR^d$ so
\bea
(f * f) (x) & = & \pi^{-2d} \int f(\alpha )f(\beta )
(D(\alpha )*D(\beta ))(x)\,d\alpha d\beta  \nonumber \\
  & = &  \pi^{-2d} \int f(\alpha )f(\beta )
  e^{2i\Theta (\alpha ,\beta )}
  e^{2i \Theta (\alpha -\beta ,x)}\,d\alpha d\beta \nonumber\\
  & = & \pi^{-2d} \int f(\alpha +\beta )f(\beta ) e^{2i\Theta (\alpha ,\beta
)}\,d\beta \,  e^{2i\Theta (\alpha ,x)}\, d\alpha.
\eea
If $f$ is, say, continuous with compact support, it follows from the last
expression that $f*f$ is the Fourier transform of a function with compact
support and therefore an entire analytic function.  In particular, its
support is $\bbR^{2d}$ unless $f=0$.
A similar argument shows that all even star powers of $f$ are analytic if $f$ has
a compact support. 

Let us now discuss odd star powers and concentrate on the third star power
for simplicity.  We find as before that
\bea
(f*f*f)(x) & = & \pi^{-3d}\int f(\alpha )f(\beta )f(\gamma )(D(\alpha)
*D(\beta) *D(\gamma) )(x)\,d\alpha d\beta d\gamma \nonumber \\
   & = & \pi^{-3d}\int  f(\alpha )f(\beta )f(\gamma ) e^{2i\Theta (\beta
,\gamma )}
e^{2i\Theta (\alpha ,\beta -\gamma )} D(x -\alpha +\beta -\gamma )\,
d\alpha d\beta d\gamma  .\nonumber\\
& &\label{3star}
\eea
from which it follows that the support of $f*f*f$ is contained in
\beq{uu}
\{ \alpha +\beta -\gamma : \alpha ,\beta ,\gamma\in {\rm supp}\, f\}.
\eeq
In particular, if the support of $f$ has diameter $R$, the support of
$f*f*f$ has diameter 
\beq{666p}
R'\leq 3R.
\eeq
In the case of, say, a quartic potential with a nonzero cubic term, i.e.,
\beq{yyy}
F(s)=V'(s)= as^3+bs^2+cs\,,
\eeq
where $b\neq 0$, we have therefore proven infinite propagation speed.   

For generic functions
$f$ we have an equality in \rf{666p}.  Here, we will not elaborate on the
detailed conditions under which this is valid.  It suffices to consider an
example which demonstrates infinite propagation speed in
the case
$b=0$ in Eq.\ \rf{yyy}.  Let
\beq{nnn}
f=D(\alpha )+D(-\alpha ),
\eeq
where $\alpha\in \bbR^{2d}$ has Euclidean norm $|\alpha |=R/2$ such that the
support of $f$ has diameter $R$.   It is easy to see that
\beq{qqq}
f*f*f= 3(D(\alpha)+D(-\alpha))+D(3\alpha)+D(-3\alpha)
\eeq
which has support diameter $R'=3R$.
Similarly, $f$ raised to the $(2n+1)$st star power
will contain delta functions at $\pm (2n+1)\alpha $.  
In order to get an example where $f$ is a smooth function, let $d^\pm_n$ be
two sequences of nonnegative smooth functions on $\bbR^{2d}$ 
with $d_n^\pm$ supported in a
ball of radius $1/n$ around $\pm \alpha$ and such that
\beq{iii}
\int d_n^\pm (x)\,dx =1.
\eeq
This ensures that
\beq{xxx}
f_n\equiv \pi^d(d_n^++d_n^-)
\eeq
is smooth and has a support diameter $R_n\leq R+2/n$.
Furthermore, 
\beq{q22}
f_n\to f=D(\alpha )+D(-\alpha )
\eeq
in the sense of distributions.  It is easily seen from \rf{3star} that
\beq{ttt}
f_n*f_n*f_n\to f*f*f
\eeq
and hence the support diameter of $f_n*f_n*f_n$ converges to $3R$ as
$n\to\infty$.   A similar argument allows one to determine the support 
of arbitrary odd star powers of compactly supported functions.

\section{Discussion}
We have proved the existence of global solutions to the initial value
problem for noncommutative nonlinear wave equations whose non-linear term is the dervative of a positive interaction potential and with
noncommutativity in the spatial directions only.  The existence
theory is quite simple and independent of the nonlinearity and the (even)
dimension of space.
 
We have shown that the speed of propagation is infinite, meaning that
the support of the solution is arbitrarily large at positive times given
that the support of the function and its first time derivative are 
compact at time $0$.  The part of the
solution which travels at infinite speed 
is proportional to the noncommutativity parameter $\theta$.

We have discussed only real solutions, but the arguments can be extended 
to cover complex wave equations such as
\beq{sss}
\pa_t^2\phi (t)+2\theta^{-1}\sum_{k=1}^d[a^*_k,[a_k,\phi (t)]]+\phi (t) F(|\phi(t)|^2 )=0\,,
\eeq
in operator form, 
where $|\phi(t)|^2=\phi(t)^*\phi(t)$ and the function $F$ is the derivative 
of a polynomial $V(s)$ that is positive for $s>0$ and vanishes linearly at $s=0$.

We can use the formalism set up in this paper to study scattering of
noncommutative waves and this will be the subject of a forthcoming
publication.  It will be interesting to see whether this scattering
theory can be used to analyse the scattering of noncommutative
solitons  and whether the results on soliton scattering at large
$\theta$ obtained in \cite{h2,sc1,sc2} can be proven rigorously.

\bigskip

\noindent
{\bf Acknowledgements} 

This work is supported in
part by MatPhySto funded by the Danish National Research Foundation and 
by TMR grant no. HPRN-CT-1999-00161.  We are indebted to the
Mittag-Leffler Institute for hospitality and to B. Birnir and R. Nest
for helpful discussions.


\begin{thebibliography}{99}

\bibitem{g1}A.~P. Polychronakos, {\it Flux tube solutions in
noncommutative gauge theories}, Phys. Lett. B {\bf 495} (2000) 407
[hep-th/0007043]

\bibitem{g2}M. Aganagic, R. Gopakumar, S. Minwalla and A.
Strominger,
{\it Unstable solitons in noncommutative gauge theory},
JHEP {\bf 0104} (2001) 001 [hep-th/0009142]

\bibitem{g3}D. Bak, {\it Exact solutions of multi-vortices and false
vacuum bubbles in noncommutative Abelian-Higgs theories},
Phys. Lett. B {\bf 495} (2000) 251
[hep-th/0008204]

\bibitem{g4}D.~J. Gross and N.~A. Nekrasov, {\it
Solitons in noncommutative gauge theory},
JHEP {\bf 0103} (2001) 044
[hep-th/0010090]

\bibitem{g5}J.~A. Harvey, P. Kraus and F. Larsen, {\it Exact
noncommutative solitons}, JHEP {\bf 0012} (2000) 024 [hep-th/0010060]

\bibitem{h1}R. Gopakumar, S. Minwalla and A. Strominger, {\it
Noncommutative solitons}, JHEP {\bf 0005} 020 (2000) [hep-th/0003160]

\bibitem{h2}  R. Gopakumar, M. Headrick, M. Spradlin, {\it
   On Noncommutative Multi-solitons},  Commun. Math. Phys. {\bf 233} (2003)
355-381 [hep-th/0103256]

\bibitem{h3} M. Spradlin and A. Volovich, {\it Noncommutative solitons
on K{\"a}hler manifolds}, JHEP {\bf 0203} (2002) 011
   [hep-th/0106180]

\bibitem{p1}B. Durhuus, T. Jonsson and R. Nest,
{\it Noncommutative scalar solitons: existence and nonexistence},
Phys. Lett. B {\bf 500}
(2001) 320 [hep-th/0011139]

\bibitem{p2}B. Durhuus, T. Jonsson and R. Nest,
{\it The existence and stability of noncommutative scalar solitons},
Commun. Math. Phys. {\bf 233} (2003) 49
[hep-th/0107121]


\bibitem{p3} B. Durhuus and T. Jonsson, {\it 
A note on noncommutative scalar multisolitons}, 
Phys. Lett. B {\bf 539} (2002) 277
[hep-th/0204096]

\bibitem{p4}P. Austing, T. Jonsson and L. Thorlacius, {\it 
Scalar Solitons on the Fuzzy Sphere},
JHEP {\bf 0210} (2002) 073 [hep-th/0206060]
                 
\bibitem{r1}A. Konechny and A. Schwarz, {\it Introduction to
(M)atrix
theory and noncommutative geometry}, Phys. Rep. {\bf 360} (2002) 354-465
[hep-th/0012145, hep-th/0107251]
 
\bibitem{r2}M.~R. Douglas and N.~A. Nekrasov, {\it
Noncommutative field theory},
Rev. Mod. Phys. {\bf 73} (2002) 977-1029
[hep-th/0106048]

\bibitem{r3}R.~J. Szabo, {\it Quantum field theory on noncommutative
spaces},  Phys. Rept. {\bf 378} (2003) 207-299
[hep-th/0109162]

\bibitem{bak}D. Bak, K. Lee, and T.-H. Park, {\it Noncommutative vertex
solitons}, Phys. Rev. D {\bf 63} (2001) 125010 [hep-th/0011099]

\bibitem{td1}A. Hashimoto and N. Itzhaki, {\it Travelling faster than
the speed of light in noncommutative geometry}, Phys. Rev. D {\bf 63}
(2001) 126004 [hep-th/0012093]

\bibitem{td2} L. Hadasz, U. Lindstrom, M. Rocek and R. von Unge, {\it
Time dependent solutions of noncommutative Chern-Simons theory coupled
to scalar fields}, UUITP-13-03, YITP-SB-03038, hep-th/0309015

\bibitem{td3}P.~A. Horvathy and P.~C. Stichel, {\it Moving vortices in
noncommutative gauge theory}, Phys. Lett. B {\bf 583} (2004) 353 
[hep-th/0311157]

\bibitem{popov1}O. Lechtenfeld and A.~D. Popov, {\it Noncommutative
multi-solitons in 2+1 dimensions}, JHEP {\bf 0111} (2001) 040
[hep-th/0106213]

\bibitem{popov2}O. Lechtenfeld and A.~D. Popov, {\it Scattering of noncommutative
solitons in 2+1 dimensions}, Phys. Lett. B {\bf 253} (2001) 178
[hep-th/0108118]

\bibitem{bieling}S. Bieling, {\it Interaction of noncommutative plane waves
in 2+1 dimensions}, J. Phys. A {\bf 35} (2002) 6281
[hep-th/0203269]

\bibitem{boost}D. Bak and K. Lee, {\it Elongation of moving
noncommutative solitons}, Phys. Lett. B {\bf 495} (2000) 231
[hep-th/0007107]

\bibitem{strauss}W. A. Strauss, {\it Nonlinear wave equations},
Regional Conference Series in Mathematics {\bf 73}, AMS, 1989.


\bibitem{reed}M. Reed, {\it Abstract nonlinear wave equations},
Lecture Notes in Math. {\bf 507}, Springer, 1976.

\bibitem{reedsimon}M. Reed and B. Simon, {\it Methods of modern
mathematical physics, II Fourier analysis, self-adjointness}, Academic
Press, 1975.

\bibitem{sc1} U. Lindstr{\"o}m, M. Rocek and R. von Unge, {\it Noncommutative
soliton scattering}, JHEP {\bf 0012} (2000) 004 [hep-th/0008108]

\bibitem{sc2}L. Hadasz, U. Lindstr{\"o}m, M. Rocek and R. von Unge, {\it
Noncommutative multisolitons: moduli spaces, quantization, finite $\theta$
effects and stability}, JHEP {\bf 0106} (2001) 040 [hep-th/0104017]

\end{thebibliography}
\end{document}